\documentclass[12pt]{aastex}
\begin{document}

\newcommand{\kms}{\,km\,s$^{-1}$}
\newcommand{\ergs}{\,erg\,s$^{-1}$}
\newcommand{\ergcm}{\,erg\,cm$^{-2}$\,s$^{-1}$}
\newcommand{\mjb}{\,mJy\,beam$^{-1}$}
\newcommand{\ppdot}{\mbox{$P$--$\dot{P}$}}
\newcommand{\pdot}{\mbox{$\dot{P}$}}
\newcommand{\edot}{\mbox{$\dot{E}$}}
\newcommand{\cxo}{\emph{Chandra}}
\newcommand{\xmm}{\emph{XMM-Newton}} 
\newcommand{\nh}{$N_\mathrm{H}$}
\newcommand{\cms}{\,cm$^{-2}$}

\title{THE CORRELATION BETWEEN DISPERSION MEASURE AND X-RAY COLUMN DENSITY
FROM RADIO PULSARS}

\author{C. He\altaffilmark{1,2}, C.-Y. Ng\altaffilmark{3,1}, and V. M.
Kaspi\altaffilmark{1} }
\altaffiltext{1}{Department of Physics, McGill University, Montreal, QC H3A 2T8, Canada}
\altaffiltext{2}{Department of Physics, University of Chicago, Chicago,
IL 60637, USA}
\altaffiltext{3}{Department of Physics, The University of Hong Kong, Pokfulam
Road, Hong Kong}
\email{ncy@bohr.physics.hku.hk}

\shorttitle{Pulsar DM-NH Correlation}
\shortauthors{He et al.}

\begin{abstract}
Pulsars are remarkable objects that emit across the entire electromagnetic
spectrum, providing a powerful probe of the interstellar medium. In this
study, we investigate the relation between dispersion measure (DM) and X-ray
absorption column density \nh\ using 68 radio pulsars detected at X-ray
energies with the \cxo\ \emph{X-ray Observatory} or \xmm. We find a best-fit
empirical linear relation of $N_\mathrm{H}\rm\;(10^{20}\,cm^{-2})=
0.30^{+0.13}_{-0.09}\;DM\;(pc\,cm^{-3})$, which corresponds to an average
ionization of $10^{+4}_{-3}$\%, confirming the ratio of one free electron per
ten neutral hydrogen atoms commonly assumed in the literature. We also compare
different \nh\ estimates and note that some \nh\ values obtained from X-ray
observations are higher than the total Galactic H{\sc i} column density along
the same line of sight, while the optical extinction generally gives the best
\nh\ predictions.
\end{abstract}


\section{INTRODUCTION}
The broadband emission of pulsars from radio frequencies to $\gamma$-rays can
be used to probe the physical conditions of the interstellar medium (ISM).
Specifically, their radio pulsations allow accurate measurements of the free
electron column density and their X-ray extinction traces the interstellar gas
along the line of sight. Radio waves travelling in the ISM are dispersed by
free electrons such that signals at lower frequencies propagate at a lower
speed and hence arrive on Earth later than those at higher frequencies. The
time delay ($\Delta t$) between two observing frequencies ($\nu_1$, $\nu_2$)
depends on the dispersion measure (DM), which is the integrated free electron
number density $n_e$ from Earth to the source at distance $d$:
\begin{equation}
\mathrm{DM} = \int_{0}^{d}{n_{e}\,\mathrm{d}l} = \frac{2\pi m_{e} c}{e^{2}} \left(
\frac{1}{\nu_1^2} - \frac{1}{\nu_2^{2}} \right)^{-1} \Delta t \ ,
\end{equation}
where $m_e$ and $e$ are electron mass and charge, respectively, and $c$ is
the speed of light. Most free electrons in our Galaxy are found in the hot
phase of the ISM, including H{\sc ii} regions ionized by UV radiation from hot
O or B type stars and the shock-heated interior of supernova remnants (SNRs).
These sources can contribute significant DM up to a few hundred parsecs per
cubic centimeter. At X-ray energies, photons are absorbed mostly by heavy
elements in the interstellar gas due to the photoelectric effect. This has a
strong energy dependence and is most prominent in the soft X-ray band. As a
result, it modifies the observed low-energy portion of the X-ray spectrum and
has to be accounted for in spectral modeling. The amount of extinction, which
is expressed in terms of the equivalent atomic hydrogen column density \nh, is
sensitive to gas and molecular clouds, which traces the warm and cold phases
of the ISM \citep[see][]{wam00}.

One natural question to ask is whether there is any correlation between DM and
\nh\ in our Galaxy. Such a correlation can reflect the physical connection
between different phases of the ISM. Also, it can provide a useful tool to
estimate one quantity from the other, help plan new observations and
determine X-ray luminosity upper limits in cases of non-detection. In the
literature, an average ionization fraction of 10\% in the ISM, i.e.\ one free
electron per 10 equivalent hydrogen atoms, has been commonly assumed in order
to infer \nh\ from DM \citep[e.g.,][]{sw88,kpg07,ghm+08,cfc+12}, but the
justification for this choice has been unclear. X-ray-emitting radio pulsars
offer a powerful diagnostic tool for a quantitative study of the correlation.
Because they are model-independent and relatively straightforward to measure
from radio timing, DM values are well determined, typically to better than a
fractional uncertainty of $10^{-3}$. However, what has made the determination
of any DM-\nh\ correlation difficult in the past is the lack of high-quality
X-ray data for \nh\ measurements. In particular, previous generations of X-ray
telescopes had poor angular resolution that precluded discerning the pulsar
emission from that of the surrounding SNRs and pulsar wind nebulae (PWNe).
Thanks to new X-ray missions such as the \cxo\ \emph{X-ray Observatory}
and \emph{XMM-Newton}, precise measurements of \nh\ have been obtained for
many pulsars in recent years, allowing a statistical study of \nh\ values for
the first time.

In this paper, we compile a list of DM and \nh\ values for 68 X-ray-emitting
radio pulsars using the latest \cxo\ and \xmm\ measurements reported in
the literature. We found a clear correlation between these two column
densities and obtained a best-fit empirical relation of $10^{+4}_{-3}$\%
ionization. In Section 2, we describe our sample selection criteria. The
statistical analysis and results are presented in Section~3, and we discuss
the implications of our results in Section~4.

\section{SAMPLE SELECTION}
We started with a list of X-ray detected radio pulsars from \citet{pcc+02},
\citet{ba02}, \citet{pkg+07}, \citet{kp08}, and \citet{kp10}, then expanded the
sample through careful literature searches for updated observational results
and recent discoveries. The latter include three magnetars that show radio
emission \citep{crh+06,crh+07,lbb+10} and over a dozen new pulsars
identified in $\gamma$-rays with the \emph{Fermi Gamma-ray Space Telescope}
and subsequently detected in follow-up radio and X-ray observations
\citep[see][]{mdc11}. Finally to complete the list, we went through the
\cxo\ and \xmm\ data archive to search for pulsar observations, and looked up 
relevant publications based on these data.

The pulsar DMs are adopted from the ATNF Pulsar
Catalog\footnote{\url{http://www.atnf.csiro.au/research/pulsar/psrcat/}}
\citep{mht+93}. They are all very well measured with negligible uncertainties
compared to those for \nh. On the other hand, it is much more difficult to
determine \nh, because this requires a strong X-ray source and good knowledge
of the intrinsic emission spectrum. The X-ray emission of pulsars is not fully
understood; commonly used models include a blackbody (BB) and a neutron-star
hydrogen atmosphere (NSA) for the thermal emission, and a power law (PL) for
the non-thermal emission. More complicated models consisting of thermal and
non-thermal components are sometimes used. To minimize any bias, we selected
the \nh\ values for our sample according to the following criteria:

\begin{enumerate}
\item We restricted our choices to those in the latest studies using the \cxo\
and \xmm\ observations, since the good angular resolution and sensitivity
of these telescopes offer high-quality spectra with minimal background
contamination. Any joint fits with other X-ray telescopes are not considered,
in order to avoid cross-calibration uncertainties.

\item We adopted only \nh\ values from actual X-ray spectral fits in which the
\nh\ is allowed to vary freely, and ignored any \nh\ inferred from DM, optical
extinction (A$_\mathrm{V}$), or total Galactic H{\sc i} column density.

\item \nh\ from the best-fit spectral model is always preferred, unless there
are physical arguments favoring another model. If different emission models
give the same goodness-of-fit and the authors do not indicate a clear
preference, we choose the simpler one. For example, we prefer a BB model
over an NSA model, since the latter requires more assumptions, including the
atmosphere composition, surface magnetic field and gravity.

\item For pulsars associated with bright PWNe, the nebular \nh\ values are
adopted if they are better constrained than those of the pulsars, because the
simple PL spectra of PWNe can reduce systematic uncertainties in spectral
modeling. \nh\ from SNRs are used in a few cases when the pulsars and PWNe are
too faint for useful \nh\ measurements.

\end{enumerate}

Our final sample contains 68 pulsars. One of them (PSR B0540$-$69) is
extragalactic and only two (PSRs J1740$-$5340 and B1821$-$24) are in globular
clusters; cluster pulsars are generally too faint for precise \nh\
measurements. The pulsar DM and \nh\ values are listed in Table~\ref{table}
and plotted in Figure~\ref{fig:dmnh}. The reported statistical uncertainties
and upper limits for \nh\ are at 90\% confidence level, i.e.\ 1.6$\sigma$.
We list in the Table the X-ray spectral models used to obtain \nh. The choice
of spectral model is clear in all cases except PSRs J1622$-$4950 and
B1757$-$24, for which both thermal and non-thermal fits are acceptable.
Nonetheless, \nh\ from different fits only varies by a factor of 2 for
J1622$-$4950 and does not change for B1757$-$24. Therefore, we conclude that
systematic bias induced by spectral models is minimal.

Table~\ref{table} also shows the pulsar Galactic coordinates ($l$, $b$) and
distances, and this information was used to calculate the vertical height
($z$) from the Galactic Plane. The coordinates are taken from the ATNF Pulsar
Catalog and distance estimates are obtained from parallax measurements, H{\sc
i} absorption measurements of the pulsars or the associated SNRs, or DM using
the NE2001 Galactic electron density model \citep{cl02}. If available,
parallax distances are always preferred since they are the most accurate. All
parallax and H{\sc i} distances are adopted from \citet{vwc+12} and references
therein, and have been corrected for the Lutz-Kelker bias, except for
PSR~J1023+0038, which has a recent parallax measurement by \citet{dab+12}. For
DM distances, we did not attempt to derive the uncertainties, but note that
the fractional uncertainties could be 25\% or larger \citep[see
e.g.,][]{cng+09}. Finally, there are exceptional cases in which previous
studies argue for different distances than the DM-estimated ones. They are
noted in the Table. The pulsar \nh\ and DM are plotted against distance in
Figures~\ref{fig:nhdist} and \ref{fig:dmdist}, respectively.

\section{ANALYSIS \& RESULTS}
\begin{figure}[ht] 
\plotone{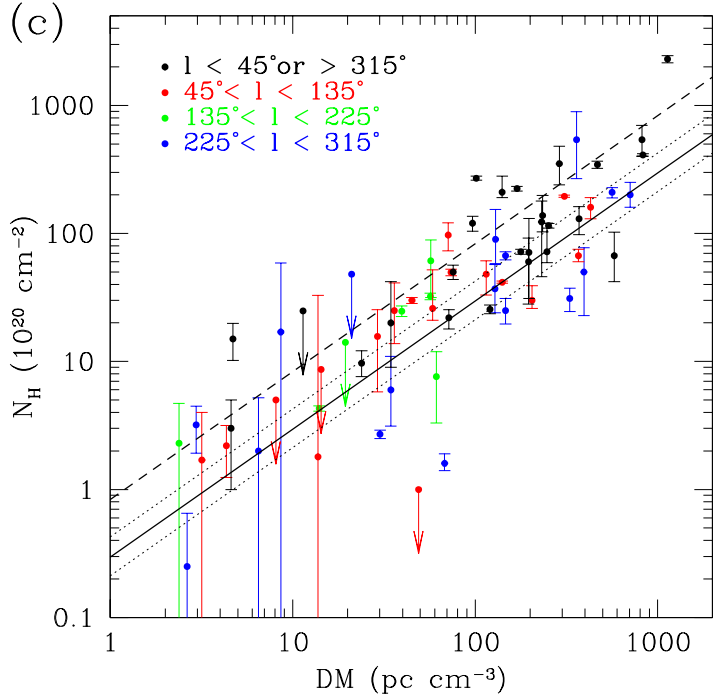}
\caption{ \nh\ versus DM for 68 pulsars. (a) Data points in gray color,
including the Vela pulsar, PSR~B0540$-$69, and \nh\ measurements with
fractional uncertainties larger than 80\% or only upper limits are not used in
the fit. The red solid and dotted lines show the best linear fit with the 90\%
confidence interval, and the blue dashed line is the same fit ignoring
measurement uncertainties. These correspond to $10^{+4}_{-3}$\% and 4\%
ionization, respectively. Uncertainties in DM are negligible. The same
plot is shown in (b) and (c) with different color schemes, indicating the
pulsar vertical distance from the Galactic Plane and their Galactic
longitudes, respectively. \label{fig:dmnh}}
\end{figure}

Figure~\ref{fig:dmnh} shows a positive correlation between the pulsar DM and
\nh\ values, with deviations ranging from a factor of a few to an order of
magnitude. There are some obvious outliers, including the Vela pulsar
(PSR~B0833$-$45), the double pulsar (PSR~J0737$-$3039), and PSR~J1747$-$2809
in the Galactic Center direction. To quantify the DM-\nh\ correlation, we
ignored pulsars with \nh\ upper limits and obtained a Pearson's correlation
coefficient of 0.72. This is significant since the one-tailed probability of
such a correlation arising by chance from unrelated variables is only
$4\times10^{-5}$. More useful is an empirical relation between these two
observables. We performed a linear fit to the data by minimizing the $\chi^2$
value. \nh\ measurements with fractional uncertainties larger than 80\% or
upper limits only (gray points in Figure~\ref{fig:dmnh}) are excluded in the
fit. We also ignored the Vela pulsar, which is located in the Gum Nebula
inside the hot and low-density Local Bubble, and PSR~B0540$-$69, which is in
the Large Magellanic Cloud (LMC), because they seem unlikely to follow the
DM-\nh\ correlation as would other Galactic sources. Only statistical
uncertainties in \nh\ are considered in the $\chi^2$-fit since uncertainties
in DM are negligible. Also, we did not attempt to model the systematic
uncertainties, but we note that the ones introduced by different photoelectric
absorption models and elemental abundances, or by cross-calibration between
telescopes are only at a few percent level \citep[see][]{wam00,tgp+11},
relatively small compared to the statistical uncertainties. Assuming \nh\ and
DM are directly proportional, the best fit gives \begin{equation} N_\mathrm{H}
\rm \;(10^{20}\,cm^{-2})=0.30^{+0.13}_{-0.09}\;DM\; (pc\,cm^{-3}) \ ,
\end{equation} corresponding to an average ionization of $10^{+4}_{-3}$\%. The
90\% confidence interval is quoted here, which is obtained from 10000
simulations via bootstrapping resampling \citep{et93}. The result is plotted
in Figure~\ref{fig:dmnh}. We also tried fitting a more general linear relation
by fitting the y-intercept as well, but found that the latter is consistent
with zero at 90\% confidence. If we ignore the measurement uncertainties in
\nh\ and perform a least squares fit, we obtain $ N_\mathrm{H}
\rm\;(10^{20}\,cm^{-2})= 0.83\;DM\; (pc\,cm^{-3})$, giving a lower average
ionization of 4\%.

\begin{figure}[ht]
\plotone{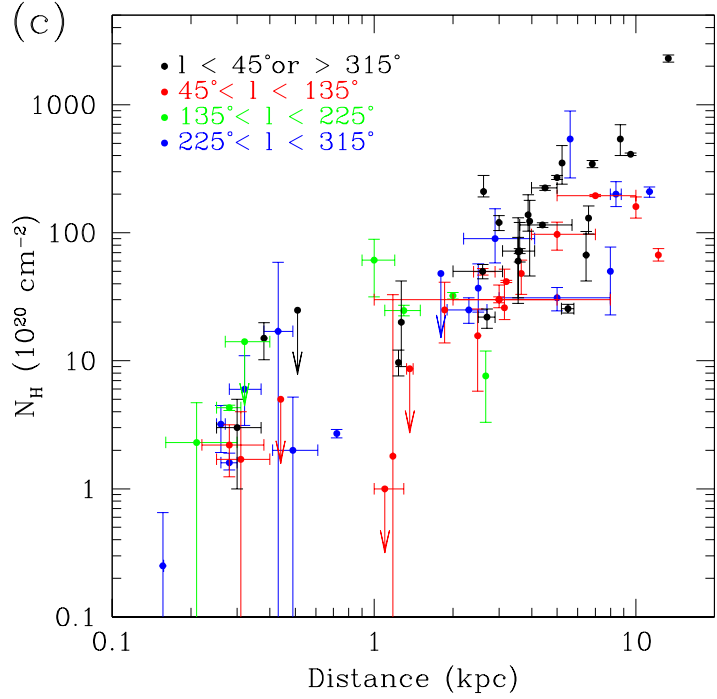} 
\caption{\nh\ versus distance for Galactic pulsars. PSR~B0540$-$69 is not
shown here. (a) The colors indicate different types of distance measurement.
We did not estimate the uncertainties for the DM distances. The same plot with
different color schemes is shown in (b) and (c), indicating the pulsar
vertical distance from the Galactic Plane and their Galactic longitudes,
respectively. \label{fig:nhdist}}
\end{figure}

\begin{figure}[ht]
\plotone{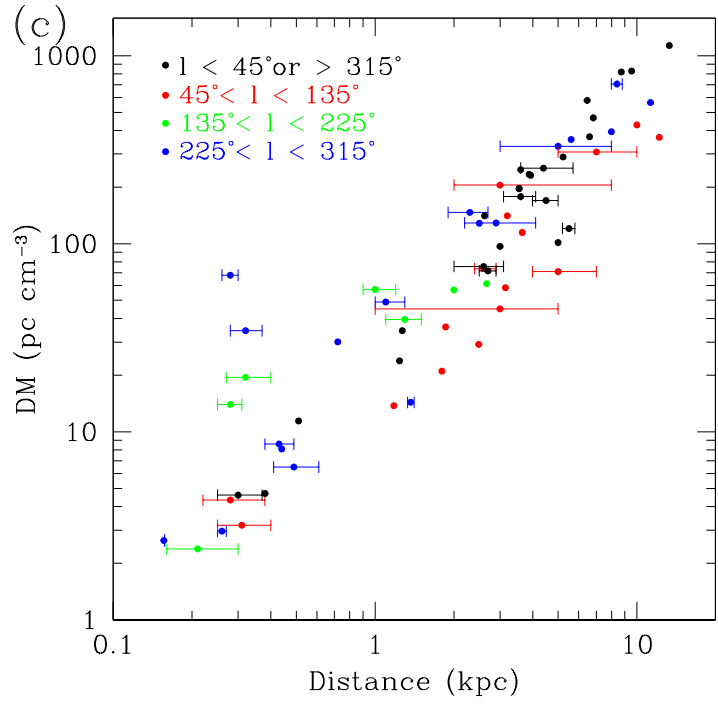}
\caption{Same as Figure~\ref{fig:nhdist}, but for DM versus distance.
Uncertainties in DM are not plotted, as they are smaller than the data points.
\label{fig:dmdist}}
\end{figure}

To check if the DM-\nh\ relation could depend on the source location in the
Galaxy, we divided the sample into groups according to their vertical height
from the plane and their Galactic longitudes. The results are shown in
Figures~\ref{fig:dmnh}(b) and \ref{fig:dmnh}(c), respectively. In the high-DM
regime, sources toward the Galactic Center direction, e.g., PSRs J1747$-$2958
and J1747$-$2809, show a hint of a larger \nh-to-DM ratio. However, the
systematic variation is less clear at lower DM and our limited sample
precludes a detailed analysis. In Figure~\ref{fig:nhdist} we plotted \nh\
against distance. This indicates a general correlation, albeit with a large
scatter. There is also a hint that for sources at a similar distance, \nh\ is
systematically larger near the Galactic Plane (Figure~\ref{fig:nhdist}(b)),
however, the dependency on Galactic longitude is less clear
(Figure~\ref{fig:nhdist}(c)). The DM variation with distance is presented in
Figure~\ref{fig:dmdist}. While this may seem to exhibit a good correlation at
large distances, we note that sources with DM-derived distances provide no new
information, only the NE2001 model prediction. In addition, there is a very
large range of DMs for nearby pulsars around 300\,pc, from $2.4\pm
0.2$\,pc\,cm$^{-3}$ for PSR~J0108$-$1431 to $68\pm1.6$\,pc\,cm$^{-3}$ for the
Vela pulsar, spanning nearly a factor of 30. Similar to \nh,
Figure~\ref{fig:dmdist}(b) also indicates a higher DM toward the Galactic
Plane.

\section{DISCUSSION}
We have investigated the DM-\nh\ connection for 68 radio pulsars detected with
\cxo\ or \xmm. We found a good correlation between these two column densities,
suggesting that free electrons in the Galaxy generally trace the interstellar
gas. That said, some \nh\ values in Figure~\ref{fig:dmnh} show significant
deviation from the best-fit line, by a factor of a few up to an order of
magnitude. This could be attributed to inhomogeneity of the ISM, possibly due
to molecular clouds, supernova remnants, or H{\sc ii} regions in the line of
sight. Such an effect is more prominent for nearby sources, since the
distribution of free electrons and interstellar gas is highly anisotropic
around the Local Bubble \citep[see][]{tc93,lwv+03}. In particular, there is
significant DM contribution from the Gum Nebula \citep{tc93}, resulting in a
wide range of DMs for pulsars within $\sim$300\,pc (e.g., the Vela pulsar and
PSR~J0737$-$3039; see Figure~\ref{fig:dmdist}). At large distances, local
fluctuations are expected to average out and the scatter of \nh\ and DM with
respect to distance likely arises from Galactic structure, such as the
disk, spiral arms, and different scale heights of various ISM components
\citep[see][]{cox05}. We have attempted to identify any systematic trends in
DM and \nh\ with respect to source location. While Figures~\ref{fig:nhdist}(b)
and \ref{fig:dmdist}(b) hint at higher \nh\ and DM toward the Galactic Plane,
more sources are needed for a quantitative comparison with the detailed
Galactic structure. Beyond our Galaxy, we note that while PSR~B0540$-$69 in
the LMC was not used in the fit, its DM-to-\nh\ ratio lies close to the
best-fit line in Figure~\ref{fig:dmnh}. This is somewhat surprising because of
the different interstellar abundances in the LMC than in our Galaxy \citep{rd92}.
We argue that this could merely be a coincidence rather than the general case.
Indeed, the LMC contributes 90\% of the \nh\ toward PSR~B0540$-$69
\citep{phs+10} but only two thirds of the DM \citep{mfl+06}.

The DM-\nh\ correlation can be used to estimate one quantity from the other,
offering a useful tool for pulsar observations. For instance, radio pulsations
have been claimed from the magnetar 4U~0142+61 with a DM of
$27\pm5$\,pc\,cm$^{-3}$ \citep{mtm10}. Given its \nh\ value of $9.6\pm0.2
\times10^{21}$\cms\ \citep{gws05}, the claimed DM seems somewhat small when
compared to other sources of similar \nh\ in Figure~\ref{fig:dmnh}. For X-ray
observations, there are many cases requiring \emph{a priori} knowledge of \nh,
including flux estimates when planning for new observations, measuring the
intrinsic spectra of faint sources, and deriving luminosity limits for
non-detection. In many previous studies, \nh\ is inferred from the DM by
assuming one free electron per ten neutral hydrogen atoms
\citep[e.g.][]{kpg07,cfc+12}. Our result directly confirms that this is a
reasonable approximation, but as a caveat, the scatter in \nh\ is typically a
factor of a few up to an order of magnitude.

\begin{figure}[ht]
\plotone{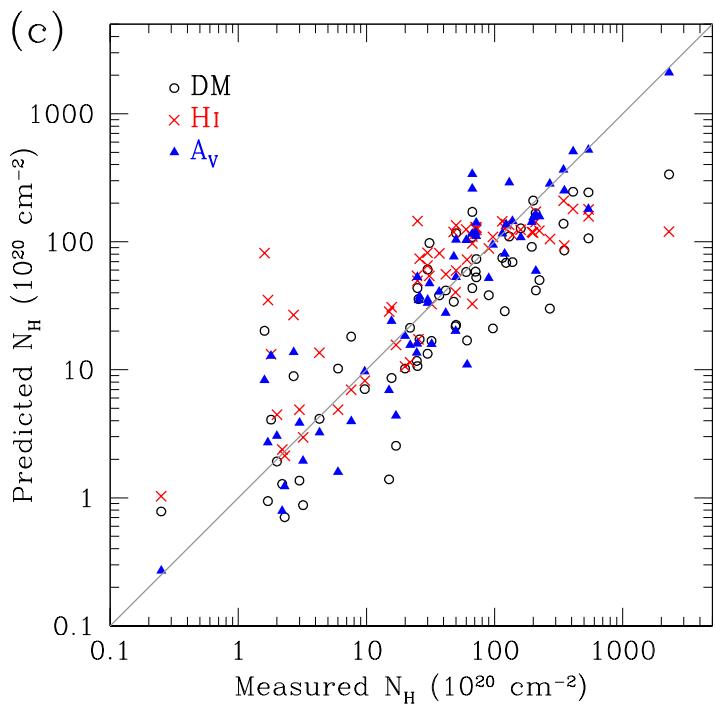}
\caption{Comparison between \nh\ estimates and measurements. The underlying
plots in (a) and (b) are the same as Figures~\ref{fig:dmnh} and
\ref{fig:nhdist}, respectively. The red crosses indicate the total
line-of-sight Galactic H{\sc i} column densities for each pulsar, given by
21\,cm radio observations \citep{kbh+05}. The blue triangles show estimates
based on A$_\mathrm{V}$ (see text). In (c), the open circles represent
predictions from our best-fit DM-\nh\ relation. \label{fig:nhest}}
\end{figure}

In addition to DM, the total Galactic H{\sc i} column density from 21-cm
radio surveys \citep[e.g.,][]{kbh+05} and A$_\mathrm{V}$ have also been used
as proxies for the X-ray absorption \citep[e.g.,][]{ozv+13}. These \nh\
estimates are plotted in Figure~\ref{fig:nhest}. It is clear that some
X-ray-inferred \nh\ values exceed the total H{\sc i} column density of the
Galaxy. As shown in the Figure, the latter saturates at $\sim10^{22}$\cms,
resulting in gross underestimates for high-DM ($\gtrsim 100$\,pc\,cm$^{-3}$)
or distant ($\gtrsim3$\,kpc) pulsars. It has been reported that at high
Galactic column densities $\gtrsim 10^{21}$\cms, which occur at low Galactic
latitudes, the X-ray absorption columns are generally larger than the H{\sc i}
columns by a factor of 1.5--3 \citep{ab99,bm06}. This agrees with our result
and indicates significant X-ray absorption due to molecular clouds rather than
neutral hydrogen atoms, hence, the H{\sc i} column may not be a good tracer
for the X-ray absorption.

A$_\mathrm{V}$, on the other hand, is caused by grains of the same heavy
elements that give rise to X-ray absorption, therefore, it highly correlates
with \nh\ \citep[e.g.,][]{ps95,go09}. Given a pulsar's position and distance,
A$_\mathrm{V}$ can be estimated from the 3D extinction maps of the Galaxy
\citep[e.g.,][]{dcl03}, and then \nh\ can be deduced from the empirical
relation \nh(\cms)=$2.21\times 10^{21}$\,A$_\mathrm{V}$ (mag) \citep{go09}. As
shown in Figure~\ref{fig:nhest}, this method seems to give the best agreement
between measured and predicted values, especially for the highest-\nh\
pulsars. It is worth noting that in some cases DMs were used to infer the
pulsar distances, which then give A$_\mathrm{V}$ and \nh. This generally
provides better results than directly employing the DM-\nh\ correlation.
We believe that this is because the A$_\mathrm{V}$ map reflects
the distribution of heavy elements in the Galaxy, whereas this crucial
information cannot be obtained from DM.

\section{CONCLUSION AND OUTLOOK}
We have compiled a list of 68 pulsar \nh\ measurements reported in the
literature using \cxo\ and \xmm\ observations, and compared the \nh\ values
with the DMs and distances. Our results show a good correlation between DM and
\nh, with a correlation coefficient of 0.72. We obtained an empirical linear
relation $N_\mathrm{H} \rm\;(10^{20}\,cm^{-2})=0.30^{+0.13}_{-0.09}\;DM\;
(pc\,cm^{-3})$, implying an average ionization of $10^{+4}_{-3}$\%. This
confirms the ratio of one free electron to ten neutral hydrogen atoms commonly
used in previous studies. Our finding provides a useful tool to estimate \nh\
from DM. We compare to other \nh\ estimates based on the neutral hydrogen
column density and A$_{\rm V}$, and find that the latter gives the best
results, while H{\sc i} and our empirical DM-\nh\ relation tend to give
underestimates in the high-\nh\ regime.

The next generation of X-ray missions, including eROSITA \citep{pab+10} and
the proposed Neutron Star Interior Composition Explorer
\citep[NICER;][]{gao12}, will significantly expand the pulsar \nh\ sample. In
addition, the foreseen Square Kilometer Array (SKA) can provide parallax
measurements of a few thousand radio pulsars \citep{stw+11}. Together these
will allow a detailed study of the DM-\nh\ relation in different parts of the
Galaxy and its connection with the Galactic structure. In addition to pulsars,
it should be possible to compile a database of \nh\ measurements for other
Galactic X-ray sources, such as stars, supernova remnants, cataclysmic
variables, stellar clusters, white dwarfs, and X-ray binaries, and compare
with their distances to build a 3D \nh\ map of our Galaxy.

\acknowledgements

We thank Oleg Kargaltsev and Slavko Bogdanov for suggesting a list of
X-ray-emitting radio pulsars, and Anne Archibald, Antoine Bouchard, and Ryan
Lynch for discussion. We acknowledge the anonymous referee for useful
suggestions. V.M.K.\ holds the Lorne Trottier Chair in Astrophysics and
Cosmology and a Canadian Research Chair in Observational Astrophysics. This
work was supported by NSERC via a Discovery Grant, by FQRNT via the Centre de
Recherche Astrophysique du Qu\'ebec, by CIFAR, and a Killam Research
Fellowship. 


\begin{deluxetable}{lccccllcc}
\tabletypesize{\scriptsize}
\tablewidth{0pt}
\tablecaption{DM, \nh, and Distances for the 68 Pulsars Used in This Study\label{table}}
\tablehead{\colhead{PSR} & \colhead{DM} &
\colhead{$N_\mathrm{H}$} & \colhead{Distance\tablenotemark{a}} & \colhead{$l$} &\colhead{$b$} &
\colhead{$z$\tablenotemark{b}} & \colhead{Model\tablenotemark{c}} & \colhead{Ref.} \\ \colhead{} &
\colhead{(pc\,cm$^{-3}$)} & \colhead{($10^{20}$\,cm$^{-2}$)} & \colhead{(kpc)} &
\colhead{(deg)} & \colhead{(deg)} & \colhead{(pc)} & \colhead{}} \startdata
J0030+0451 & 4.333$\pm$0.001 & 2.2$\pm$1.0 & $0.28^{+0.10}_{-0.06}$$^{\rm p}$
& 113.1 & $-57.6$ & $-236$ & NSA$\times$3 & 1 \\
J0108$-$1431 & $2.4\pm0.2$ & $2\pm2$ & $0.21^{+0.09}_{-0.05}$$^{\rm p}$ & 140.9 & $-76.8$ & $-205$ & BB & 2\\
B0136+57 & 73.779$\pm$0.006 & $50\pm3$ & $2.6^{+0.3}_{-0.2}$$^{\rm p}$ & 129.2 & $-4.0$ & $-183$ & PL & 3 \\ %
J0205+6449 & 140.7$\pm$0.3 & 41.6$^{+0.8}_{-0.7}$ & 3.2$^{\rm o}$ & 130.7 &
+3.1 & +172 & PL+RS (SNR) & 4, 5\\
J0218+4232 & 61.252$\pm$0.005 & $8\pm4$ & $2.67$$^{\rm d}$ & 139.5 & $-17.5$ & $-804$ & PL & 6 \\
B0355+54 & 57.1420$\pm$0.0003 & $60\pm30$ & $1.0^{+0.2}_{-0.1}$$^{\rm p}$ & 148.2 & +0.8 & +14 & PL (PWN) & 7\\
J0437$-$4715 & 2.64476$\pm$0.00007 & $0.25^{+0.40}_{-0.24}$ & $0.156\pm0.001$$^{\rm p}$ & 253.4 & $-42.0$ & $-104$ & PL+NSA$\times$2 & 8 \\
B0531+21 (Crab) & 56.791$\pm$0.001 & 32$\pm$2 & 2.00$^{\rm o}$ & 184.6 & $-5.8$ & $-202$ & PL & 9, 10\\
J0538+2817 & 39.570$\pm$0.001 & $25\pm2$ & $1.3\pm{0.2}$$^{\rm p}$ & 179.7 & $-1.7$ & $-38$ & BB & 11 \\
B0540$-$69 & $146.6\pm0.2$ & $67\pm5$ & 50$^{\rm o}$ & 279.7 & $-31.5$ & $-26137$ & PL (SNR) & 12, 13\\
B0628$-$28 & 34.468$\pm$0.017 & $6^{+5}_{-3}$ & $0.32^{+0.05}_{-0.04}$$^{\rm p}$ & 237.0 & $-$16.8 & $-$92.3 & PL & 14\\
B0656+14 & 13.977$\pm$0.013 & 4.3$\pm$0.2 & $0.28\pm0.03$$^{\rm p}$ & 201.1 & +8.3 & +40 & PL+BB$\times$2 & 15 \\
J0737$-$3039 & 48.920$\pm$0.005 & $<1$ & $1.1^{+0.2}_{-0.1}$$^{\rm p}$ & 245.2 & $-$4.5 & $-$86 & BB$\times$2 & 16 \\%
B0823+26 & 19.454$\pm$0.004 & $<14$ & $0.32^{+0.08}_{-0.05}$$^{\rm p}$ & 197.0 & +31.7 & +168 & PL & 17\\ %
B0833$-$45 (Vela) & 67.99$\pm$0.01 & $1.6^{+0.3}_{-0.2}$ & $0.28\pm0.02$$^{\rm p}$ & 263.6 & $-2.8$ & $-14$ & PL (PWN) & 18 \\ 
B0950+08 & 2.958$\pm$0.003 & 3.2$\pm$1.3 & 0.261$\pm$0.005$^{\rm p}$ & 228.9 & +43.7 & +180 & PL+BB &19\\
J1016$-$5857 & 394.2$\pm$0.2 & 50$\pm$30 & 8.00$^{\rm d}$ & 284.1 & $-1.9$ & $-263$ & PL & 20\\
J1023+0038 & 14.325$\pm$0.010 & $<9$ & 1.37$\pm$0.04$^{\rm p}$ & 243.5 & +45.8 & +980 & PL+NSA & 21 \\%
J1024$-$0719 & 6.48520$\pm$0.00008 & $2^{+3}_{-2}$ & $0.49^{+0.12}_{-0.08}$$^{\rm p}$ & 251.7 & +40.5 & +318 & BB & 22 \\
B1046$-$58 & 129.1$\pm$0.2 & 90$^{+60}_{-30}$ & $2.9^{+1.2}_{-0.7}$$^{\rm h}$ & 287.4 & +0.6 & +29 & PL & 23 \\
B1055$-$52 & 30.1$\pm$0.5 & $2.7\pm0.2$ & 0.72$^{\rm d}$ & 286.0 & +6.7 & +84 & PL+BB$\times$2 & 15 \\
J1119$-$6127 & 707.4$\pm$1.3 & $200^{+50}_{-40}$ & 8.4$\pm$0.4$^{\rm o}$ & 292.2 & $-0.5$ & $-78.7$ & PL+BB & 24, 25 \\
J1124$-$5916 & 330$\pm$2 & 31$\pm$6 & 5$^{+3}_{-2}$$^{\rm h}$ & 292.0 & +1.8 & +153 & PL & 26 \\
J1231$-$1411 & 8.090$\pm$0.001 & $<5$ & 0.44$^{\rm d}$ & 295.5 & +48.4 & +329 & NSA+PL & 27 \\ 
B1259$-$63 & 146.72$\pm$0.03 & $25^{+6}_{-5}$ & 2.3$\pm$0.4$^{\rm o}$ & 304.2 & $-$1.0 & $-$40 & PL & 28, 29\\ 
J1357$-$6429 &128.5$\pm$0.7 & 37$^{+20}_{-13}$ & 2.50$^{\rm d}$ & 309.9 & $-$2.5 & $-$110 & PL (PWN) & 30 \\
J1400$-$6325 & 563$\pm$4 & $209\pm20$ & 11.27$^{\rm d}$ & 310.6 & $-$1.6 & $-$313 & PL (PWN) &  31\\
J1420$-$6048 & 358.8$\pm$0.2 & $540^{+350}_{-270}$ & 5.61$^{\rm d}$ & 313.5 & +0.2 & +22 & PL (PWN) & 32 \\
B1451$-$68 & 8.6$\pm$0.2 & 17$^{+40}_{-17}$ & $0.43^{+0.06}_{-0.05}$$^{\rm p}$ & 313.9 & $-$8.5 & $-64$ & PL+BB & 33\\
J1509$-$5850 & 140.6$\pm$0.8 & 210$^{+70}_{-20}$ & 2.62$^{\rm d}$ & 320.0 & $-$0.6 & $-$28 & PL (PWN) &34 \\ 
B1509$-$58 & 252.5$\pm$0.3 & 115$\pm5$\tablenotemark{d} & 4.4$^{+1.3}_{-0.8}$$^{\rm h}$ & 320.3 & $-$1.2 & $-$89 & PL (PWN) & 35 \\
J1550$-$5418 & 830$\pm$50 & 410$\pm$10 & 9.55$^{\rm d}$ & 327.2 & $-$0.1 & $-$22 & PL+BB & 36 \\%
J1614$-$2230 & 34.4865$\pm$0.0001 & 20$^{+22}_{-11}$ & 1.27$^{\rm d}$ & 352.6
& +20.2 & +438 & BB$\times$2 & 37 \\%
J1617$-$5055 & 467$\pm$5 & 345$\pm20$ & 6.82$^{\rm d}$ & 332.5 & $-$0.3 & $-$33 & PL & 38 \\ 
J1622$-$4950 & 820$\pm$30 & 540$^{+160}_{-140}$ & 8.73$^{\rm d}$ & 333.8 & $-$0.1 & $-$16 & BB & 39 \\%
B1706$-$44 & 75.69$\pm$0.05 & 50$\pm6$ & 2.6$^{+0.5}_{-0.6}$$^{\rm h}$ & 343.1 & $-$2.7 & $-$122 & PL (PWN) & 40\\
J1718$-$3718 & 371.1$\pm$1.7 & 130$\pm30$ & 6.6$^{\rm d}$ & 349.8 & +0.2 & +25 & BB & 41\\ %
J1718$-$3825 & 247.4$\pm$0.3 & 72$^{+50}_{-13}$ & 3.6$^{\rm d}$ & 349.0 & $-$0.4 & $-$27 & PL (PWN) & 42 \\
J1734$-$3333 & 578$\pm$9 & $70^{+40}_{-30}$ & 6.46$^{\rm d}$ & 354.8 & $-$0.4 & $-$49 & BB & 43 \\ %
J1740$-$5340 & 71.8$\pm$0.2 & $22\pm4$ & 2.7$\pm$0.2$^{\rm o}$ & 338.2 & $-$12.0 & $-$560 & PL & 44, 45\\ %
J1740+1000 & 23.85$\pm$0.05 & $10\pm2$ & 1.24$^{\rm d}$ & 34.0 & +20.3 & +430 & PL+BB & 46\\ %
J1741$-$2054 & 4.7$\pm$0.1 & 15$\pm5$ & 0.38$^{\rm d}$ & 6.4 & +4.9 & +33 & PL (PWN) & 47 \\
J1747$-$2809 & 1133$\pm$3 & 2300$\pm$150 & 13.31$^{\rm d}$ & 0.9 & +0.1 & +18 & PL (PWN) & 48\\
J1747$-$2958 & 101.5$\pm$1.6 & 270$\pm$10 & 5$^{\rm o}$ & 359.3 & $-$0.8 & $-$73 & PL (PWN) & 49, 50 \\
B1757$-$24 & 289$\pm$10 & 350$^{+130}_{-110}$ & 5.22$^{\rm d}$ & 5.3 & $-$0.9 & $-$80 & PL & 51 \\
B1800$-$21 & 233.99$\pm$0.05 & 138$^{+60}_{-35}$& 3.88$^{\rm d}$ & 8.4 & +0.2 & +10 & PL (PWN) & 52\\
J1809$-$1917 & 197.1$\pm$0.4& 71$^{+60}_{-40}$ & 3.55$^{\rm d}$ & 11.1 & +0.1 & +5 & PL & 53\\
J1809$-$1943 & 178$\pm$5 & 72$\pm3$ & $3.6\pm0.5$$^{\rm h}$ & 10.7 & $-$0.2 & $-$10 & BB$\times$3 & 54 \\%
J1819$-$1458 & 196.0$\pm$0.4 & 60$\pm30$ & 3.55$^{\rm d}$ & 16.0 & +0.1 & +5 & BB & 55 \\%
B1821$-$24 & 120.502$\pm$0.002 & $26\pm2$ & 5.5$\pm$0.3$^{\rm o}$ & 7.8 & $-$5.6 & $-$535 & PL & 56, 57\\
B1823$-$13 & 231$\pm$1 & $120^{+60}_{-80}$ & 3.93$^{\rm d}$ & 18.0 & $-$0.7 & $-$47 & PL (PWN) & 58\\
J1833$-$1034 & 169.5$\pm$0.1 & $224^{+9}_{-10}$ & $4.5\pm0.5^{\rm h}$ & 21.5 & $-$0.9 & $-$70 & PL & 59 \\%
B1853+01 & 96.74$\pm$0.12 & 120$\pm16$ & $3^{\rm o}$ & 34.6 & $-0.5$ & $-26$ & VNEI$\times$2 (SNR) & 60, 61\\
J1930+1852 & 308$\pm$4 & 195$\pm$4 & $7^{+3}_{-2}$$^{\rm h}$ & 54.1 & +0.3 & +32 & PL (PWN) & 62\\
B1929+10 & 3.180$\pm$0.004 & $1.7^{+2.3}_{-1.7}$ & $0.31^{+0.09}_{-0.06}$$^{\rm p}$ & 47.4 & $-3.9$ & $-21$ & PL+BB& 63\\
B1937+21 & 71.0398$\pm$0.0002 & $97\pm24$ & $5^{+2}_{-1}$$^{\rm p}$ & 57.5 &
$-0.3$ & $-25$ & PL & 64\\
B1951+32 & 45.006$\pm$0.0190 & $30\pm2$ & $3\pm2^{\rm h}$ & 68.8 & +2.8 & +148 & PL & 65 \\
B1957+20 & 29.1168$\pm$0.0007 & $16\pm10$ & 2.49$^{\rm d}$ & 59.2 & $-4.7$ & $-204$ & PL & 66\\
J2021+3651 & 368$\pm$1 & 67$^{+8}_{-7}$ & 12.19$^{\rm d}$ & 75.2 & +0.1 & +24 & PL & 67\\ 
J2022+3842 & 429.1$\pm$0.5 & 160$\pm$30 & 10$^{\rm o}$ & 76.9 & +1.0 & +168 & PL & 68\\%
J2032+4127 & 114.8$\pm$0.1 & $48^{+13}_{-15}$ & 3.65$^{\rm d}$ & 80.2 & +1.0 & +66 & PL &69 \\%
J2043+2740 & 21.0$\pm$0.1 & $<50$ & $1.8^{\rm d}$ & 70.6 & $-9.2$ & $-286$ & BB & 17 \\%
J2124$-$3358 & 4.601$\pm$0.003 & 3$\pm$2 & $0.30^{+0.07}_{-0.05}$$^{\rm p}$ & 10.9 & $-45.4$ & $-214$ &PL+BB & 22 \\ 
B2224+65 & 36.079$\pm$0.009 & 25$^{+16}_{-11}$ & 1.86$^{\rm d}$ & 108.6 & +6.8 & +222 & PL & 70\\%
J2229+6114 & 204.97$\pm$0.02 & 30$^{+9}_{-4}$ & $3^{+5}_{-1}$$^{\rm o}$ & 106.6 & +2.9 & +154 & PL & 71, 72 \\%
J2241$-$5236 & 11.41085$\pm$0.00003 & $<25$ & 0.51$^{\rm d}$ & 337.5 & $-$54.9 & $-417$ & PL & 69 \\%
J2302+4442 & 13.762$\pm$0.006 & $2^{+31}_{-2}$ & 1.18$^{\rm d}$ & 103.4 & $-$14.0 & $-286$ & NSA & 73 \\%
B2334+61 & 58.410$\pm$0.015 & 26$^{+26}_{-5}$ & 3.15$^{\rm d}$ & 114.3 & +0.2 & +13 & BB & 74
\enddata

\tablenotetext{a}{Distance estimates are either parallax or H{\sc i}
absorption measurements adopted from \citet{vwc+12} and \citet{dab+12} or from
the DM using the NE2001 model \citep{cl02}. They are denoted by the letters
p, h, and d, respectively. We did not attempt to quantify the uncertainties in
the DM distances. The designation ``o'' indicates distance estimates based on
other arguments; see the references for details.}

\tablenotetext{b}{Vertical distance from the Galactic Plane calculated using
the source distance and Galactic latitude $b$.}

\tablenotetext{c}{Spectral models used to obtain \nh: blackbody (BB), power law
(PL), and neutron star atmosphere (NSA). \nh\ values determined from the
associated SNRs or PWNe are noted. The SNR spectra were fitted with 
Raymond-Smith (RS) and non-equilibrium ionization (VNEI) models.}

\tablenotetext{d}{\citet{sbd+10} reported $N_\mathrm{H}=1.15\times10^{21}$\cms,
but mentioned that it is consistent with the previous result from
\citet{gak+02}, which gave $N_\mathrm{H}=9.5\times10^{21}$\cms. Therefore, the
former is assumed to be a typographical error and we adopt the value
$1.15\times10^{22}$\cms.}.

\tablecomments{Uncertainties and upper limits on \nh\ are all scaled to the
90\% confidence level, i.e.\ $1.6\sigma$.}

\tablerefs{(1) \citealt{bg09}; (2) \citealt{pap+12}; (3) \citealt{mcm+11}; (4)
\citealt{ghn07}; (5) \citealt{rgk+93}; (6) \citealt{wob04}; (7)
\citealt{to07}; (8) \citealt{dkp+12}; (9) \citealt{wty+11}; (10)
\citealt{tri73}; (11) \citealt{nrb+07}; (12) \citealt{phs+10}; (13)
\citealt{fmg+01}; (14) \citealt{bjk+05}; (15) \citealt{dcm+05}; (16)
\citealt{prm+08}; (17) \citealt{bwt+04}; (18) \citealt{lsd08}; (19)
\citealt{zp04}; (20) \citealt{cgg+04}; (21) \citealt{bah+11}; (22)
\citealt{zav06}; (23) \citealt{gkp+06}; (24) \citealt{nkh+12}; (25)
\citealt{cmc04}; (26) \citealt{hsp+03}; (27) \citealt{rrc+11}; (28)
\citealt{pck11}; (29) \citealt{nrh+11}; (30) \citealt{cpk+12}; (31)
\citealt{rmg+10}; (32) \citealt{nrr05}; (33) \citealt{ppm+12}; (34)
\citealt{kmp+08}; (35) \citealt{sbd+10}; (36) \citealt{nkd+11}; (37)
\citealt{pwb+12}; (38) \citealt{kpw09}; (39) \citealt{ags+12}; (40)
\citealt{rnd+05}; (41) \citealt{zkm+11}; (42) \citealt{hfc+07}; (43)
\citealt{ozv+13}; (44) \citealt{bvh+10}; (45) \citealt{rg98}; (46)
\citealt{kdm+12}; (47) \citealt{rsc+10} (48) \citealt{hse+12}; (49)
\citealt{gvc+04}; (50) \citealt{gvc+04}; (51) \citealt{kgg+01}; (52)
\citealt{kpg07}; (53) \citealt{kp07}; (54) \citealt{bid+09}; (55)
\citealt{rmg+09}; (56) \citealt{bvs+11}; (57) \citealt{har96}; (58)
\citealt{pkb08}; (59) \citealt{ms10}; (60) \citealt{skp04}; (61)
\citealt{cwb+96}; (62) \citealt{tsr+10}; (63) \citealt{mpg08}; (64) Ng et al.\
2013, in preparation; (65) \citealt{lll05}; (66) \citealt{gjv+12}; (67)
\citealt{vrn08}; (68) \citealt{agr+11}; (69) \citealt{mar12}; (70)
\citealt{hht+12}; (71) \citealt{mdc11}; (72) \citealt{hgl+01}; (73)
\citealt{cgj+11}; (74) \citealt{mzc+06}}

\end{deluxetable}

\epsscale{0.6} 
\begin{figure}[ht] 
\plotone{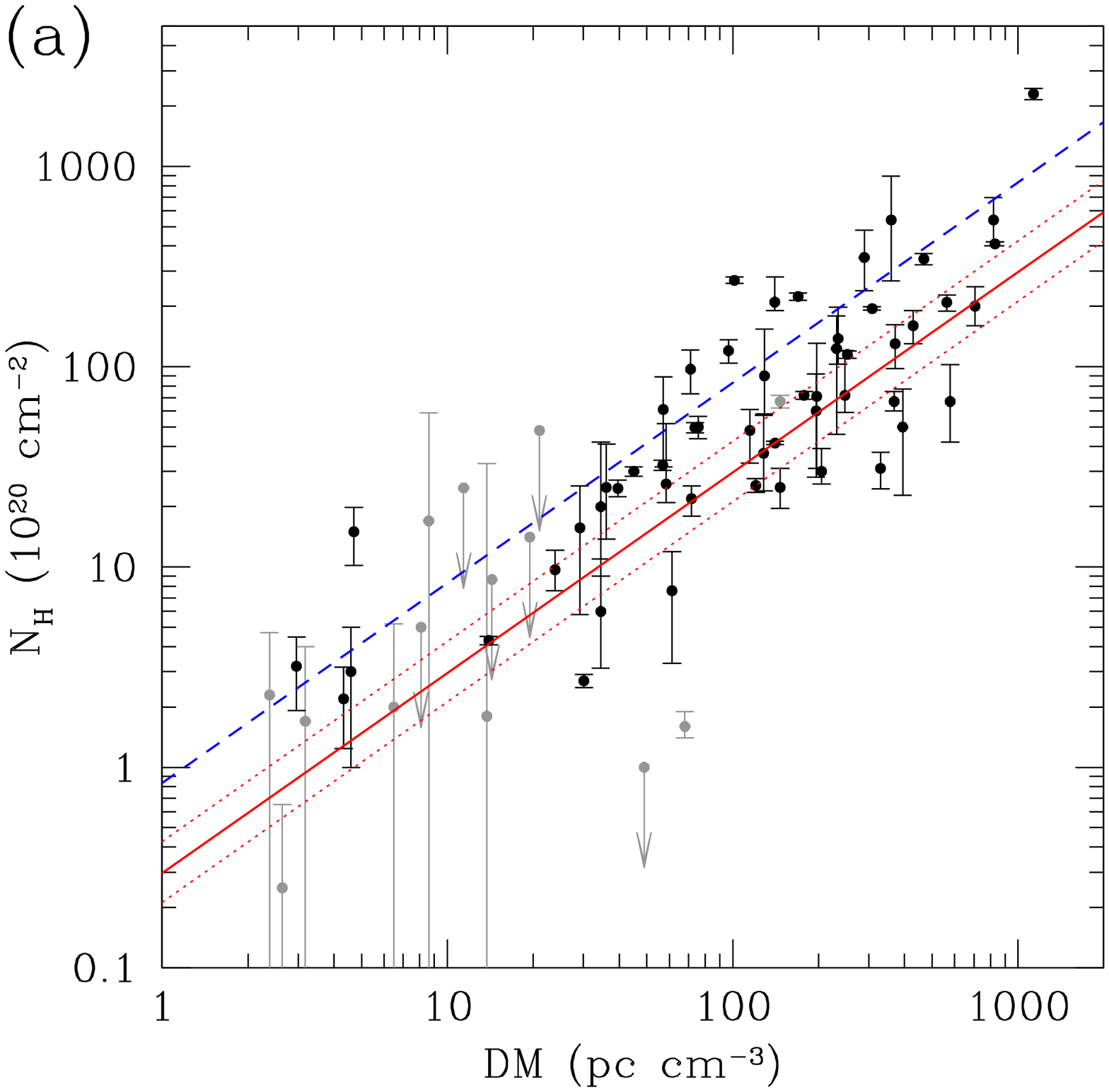}
\plotone{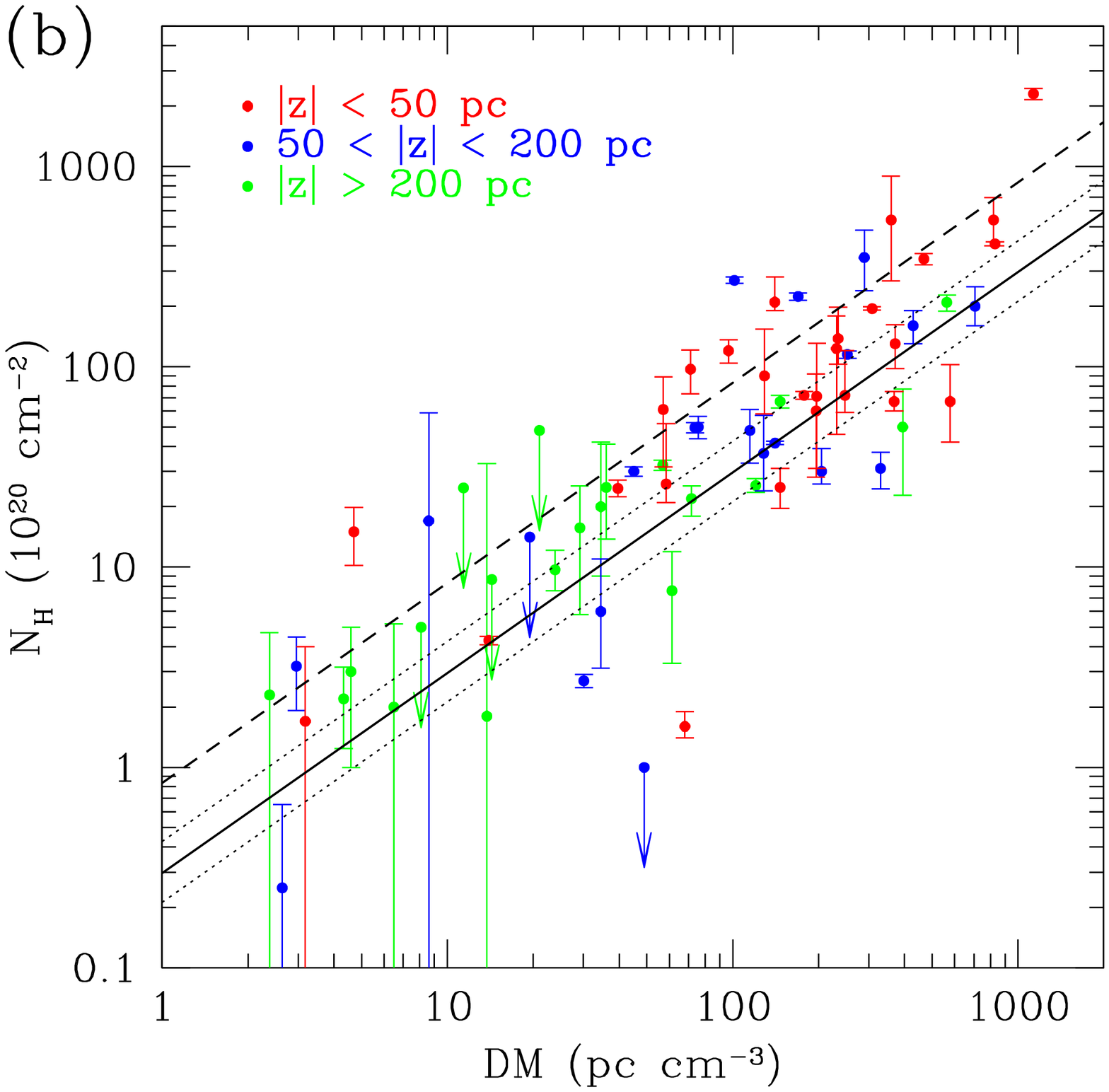}
\end{figure}
\clearpage

\clearpage

\begin{figure}[ht] 
\plotone{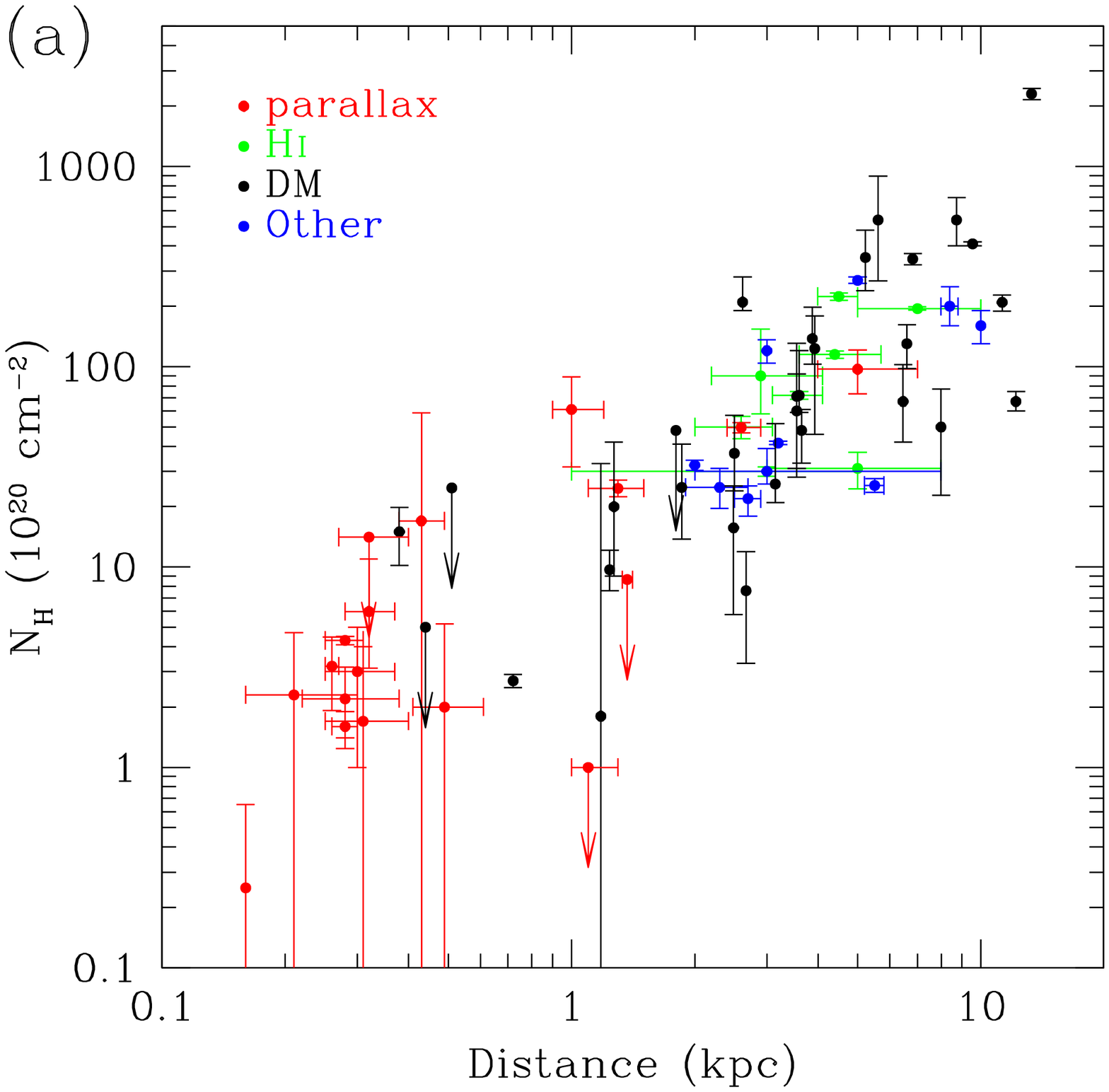}
\plotone{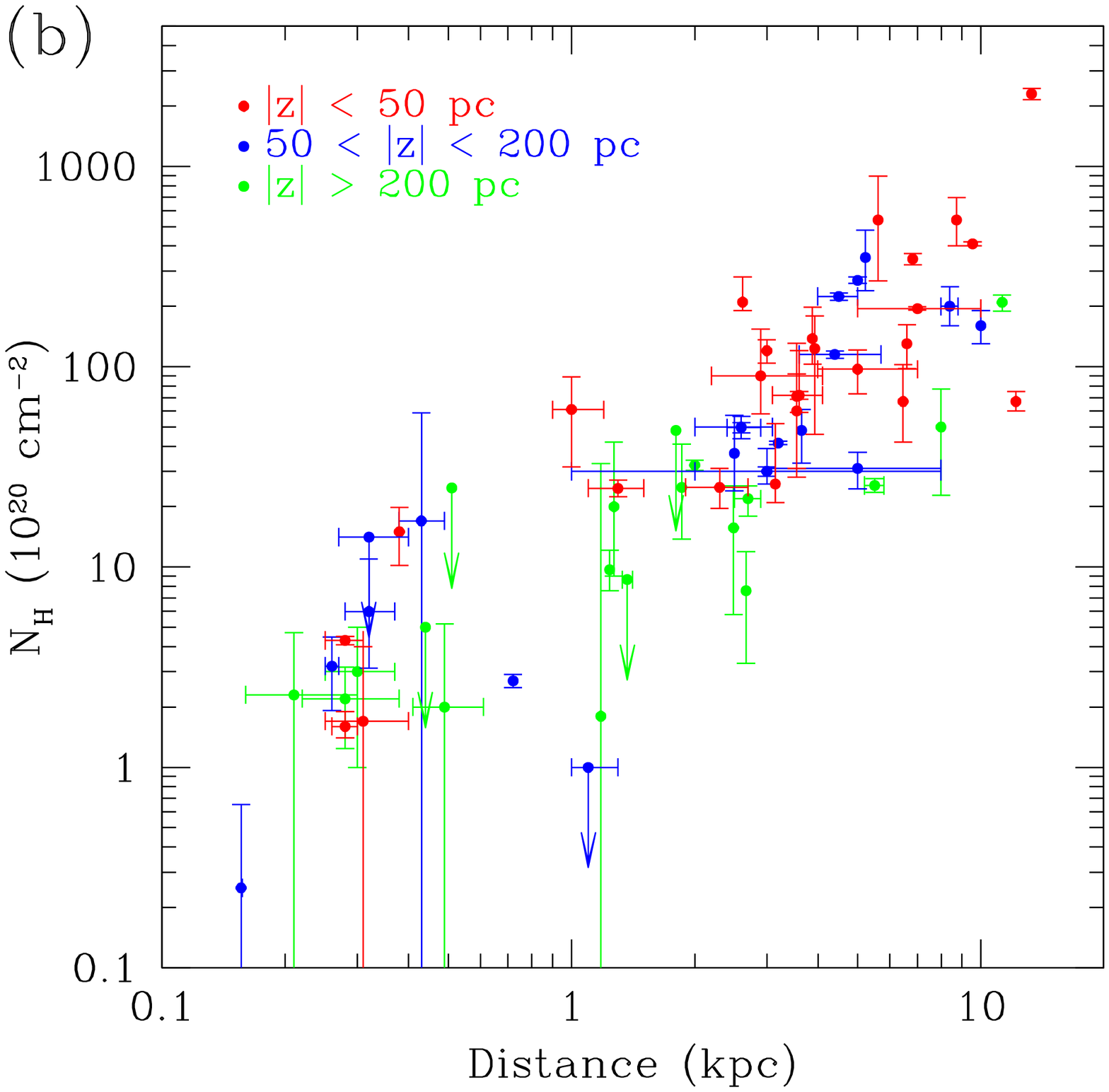}
\end{figure}
\clearpage

\clearpage

\begin{figure}[ht] 
\plotone{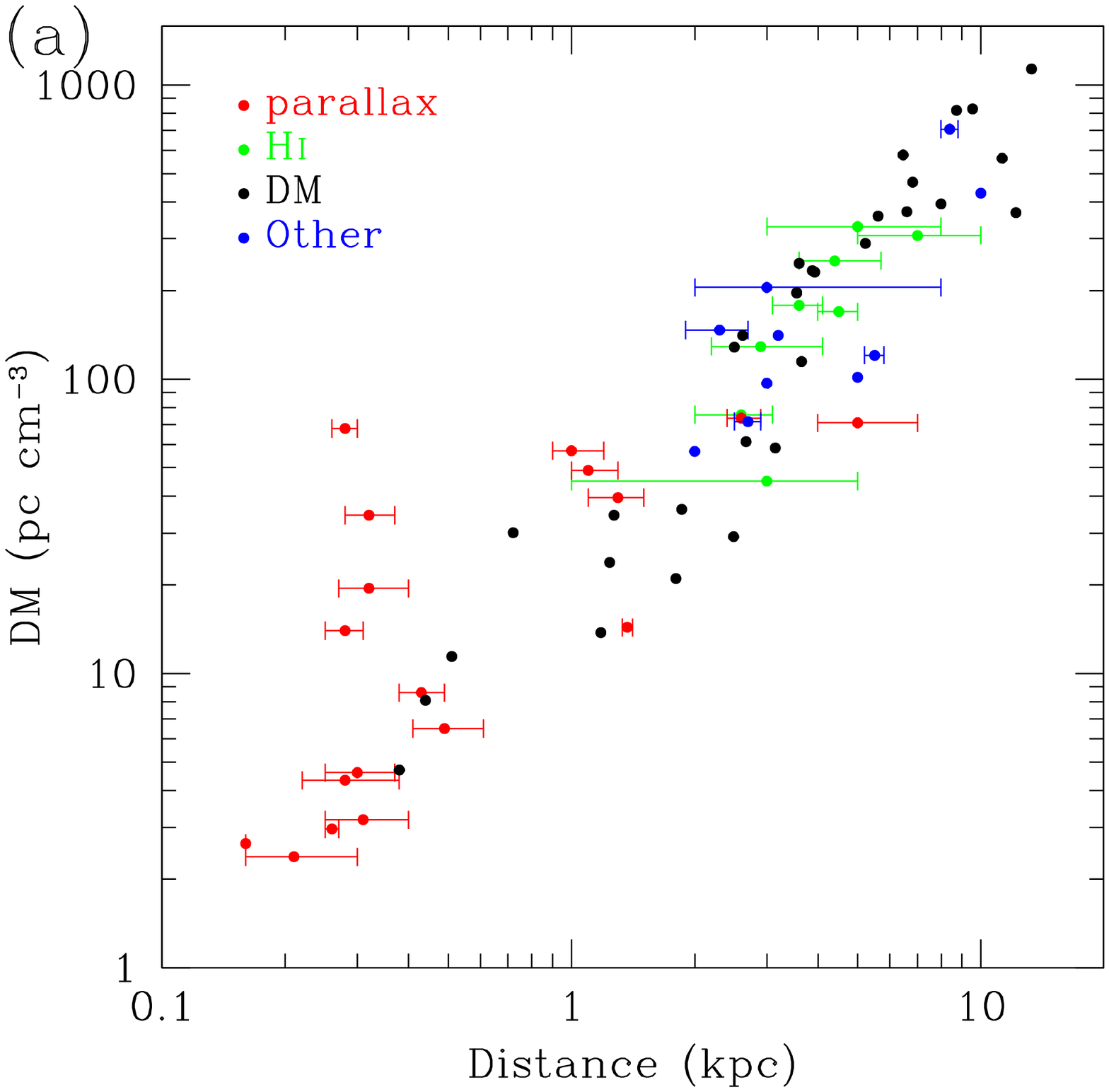}
\plotone{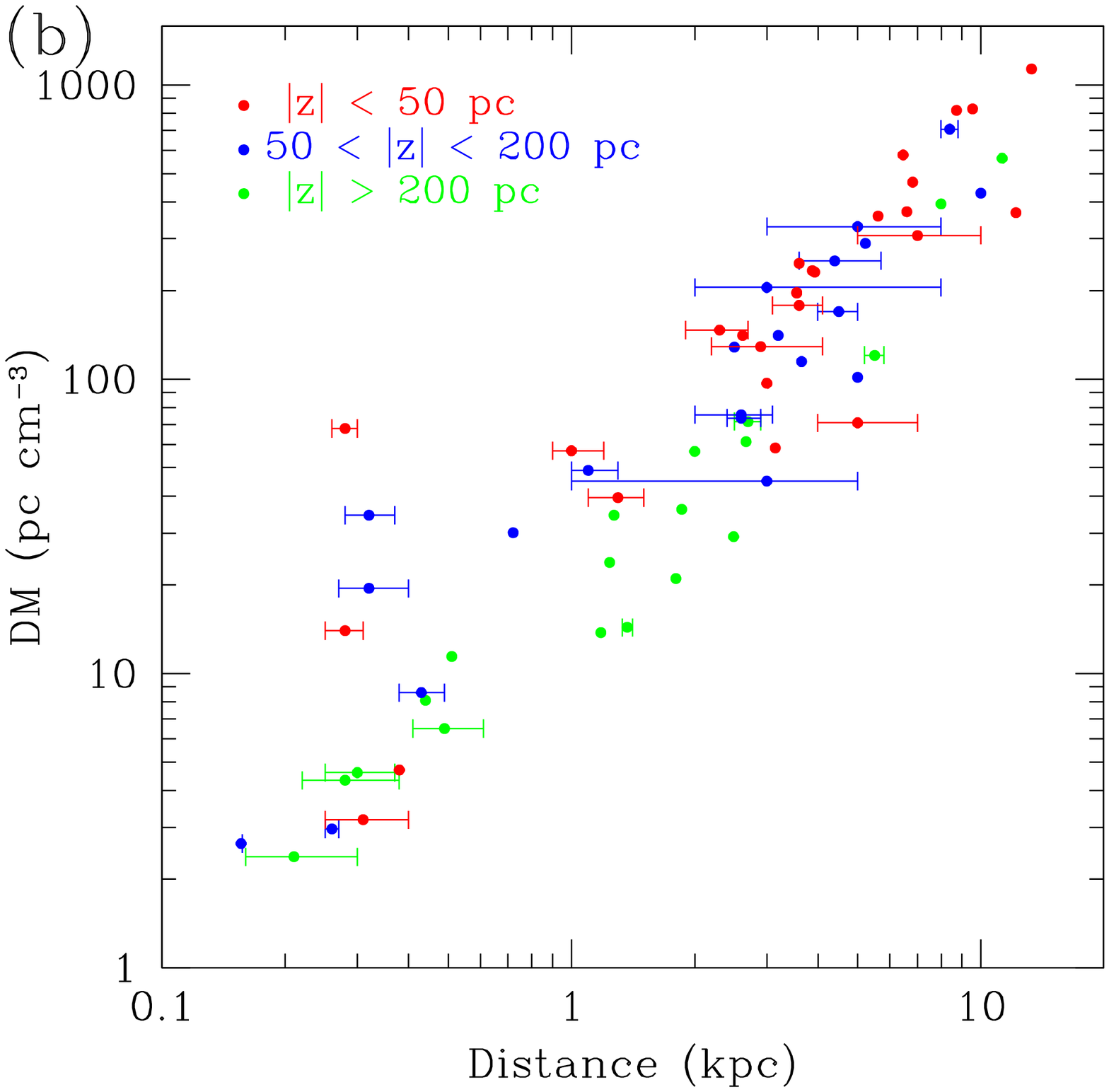}
\end{figure}
\clearpage

\begin{figure}[ht]
\plotone{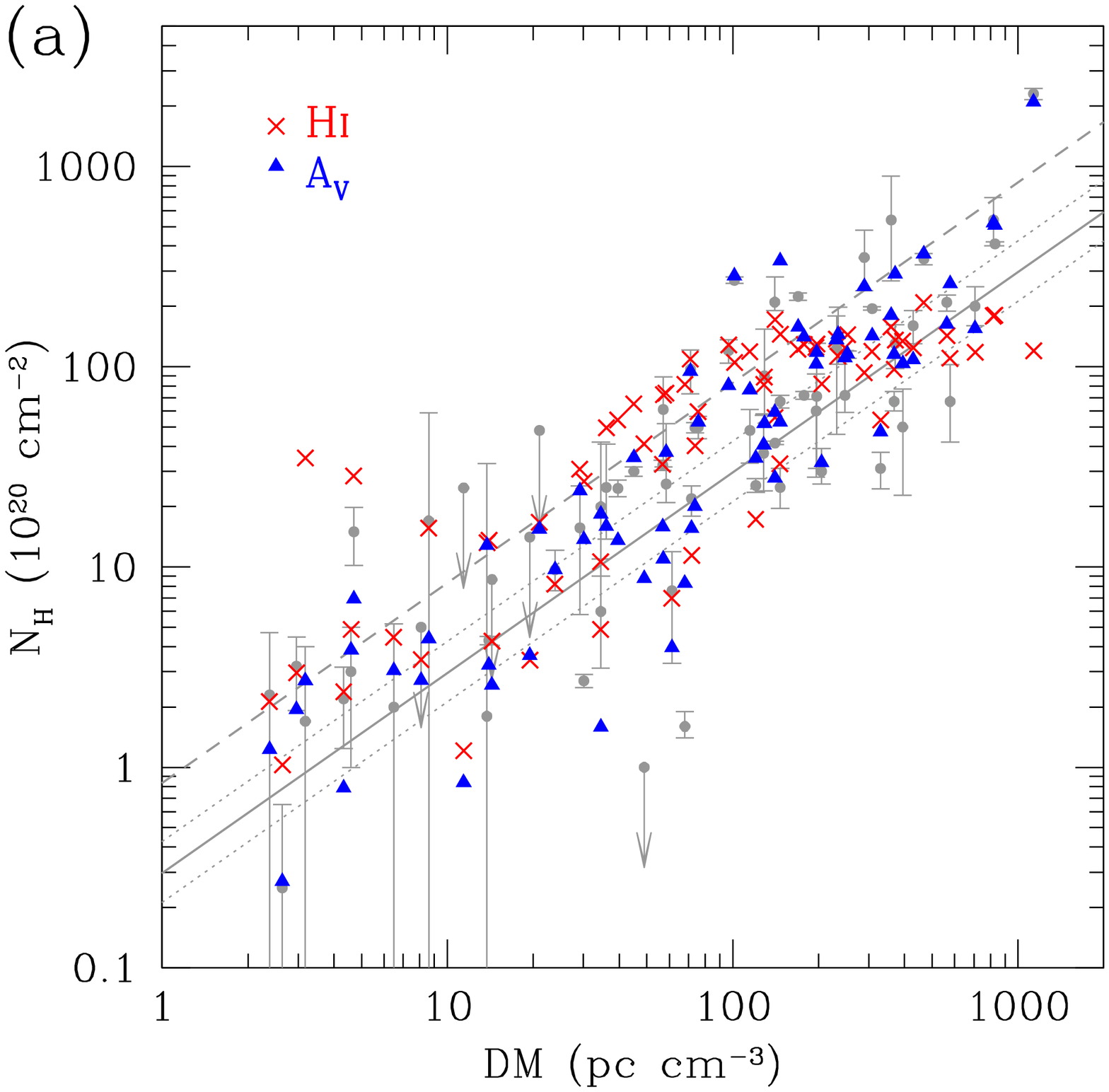}
\plotone{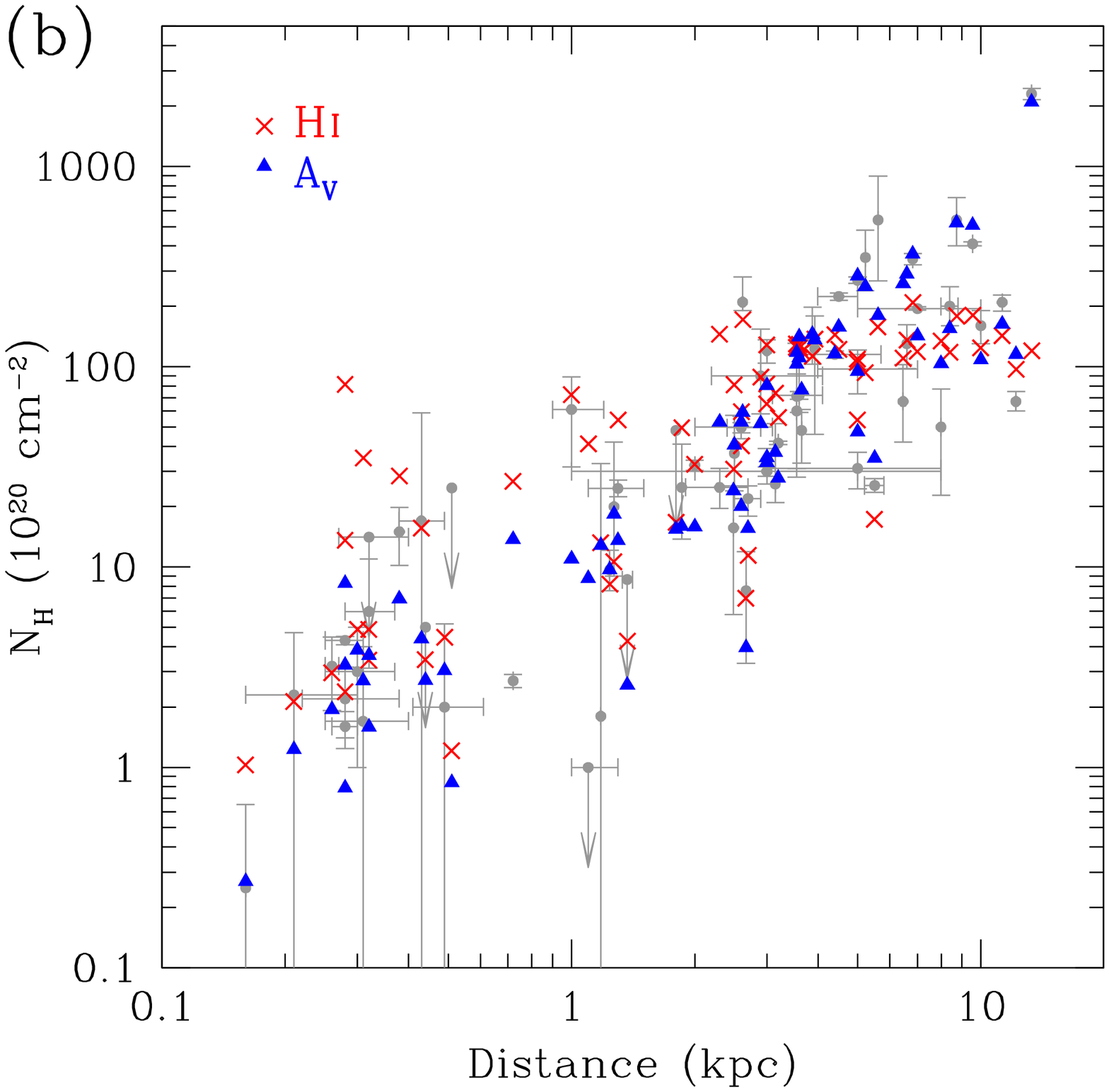}
\end{figure}
\clearpage

\end{document}